\documentclass[
aip,
 amsmath,amssymb,
 reprint
]{revtex4-2}

\usepackage{graphicx}
\usepackage{dcolumn}
\usepackage{bm}
\usepackage{array} 
\usepackage{tabularx}

\usepackage[utf8]{inputenc}
\usepackage[T1]{fontenc}
\usepackage{amsmath}
\usepackage{hyperref}
\usepackage{caption}
\usepackage{subcaption}

\makeatletter
\def\@email#1#2{%
 \endgroup
 \patchcmd{\titleblock@produce}
  {\frontmatter@RRAPformat}
{\frontmatter@RRAPformat{\produce@RRAP{*\href{mailto:#2}{#2}}}\frontmatter@RRAPformat}
  {}{}
}%

\begin{document}

\preprint{AIP/123-QED}
\title{Decoupling of Spin-Orbit Torque Components in Py/W Bilayers unveiled through variation of W-resistivity}

\author{Abu Bakkar Miah}
\affiliation{Department of Physical Sciences,
 Indian Institute of Science Education and Research Kolkata, Mohanpur, West Bengal,741246, INDIA.}
\author{Harekrishna Bhunia}
 \affiliation{Department of Physical Sciences,
 Indian Institute of Science Education and Research Kolkata, Mohanpur, West Bengal,741246, INDIA.}
 \author{Dhananjaya Mahapatra}
 \affiliation{Department of Physical Sciences,
 Indian Institute of Science Education and Research Kolkata, Mohanpur, West Bengal,741246, INDIA.}
 \author{Soumik Aon}
 \affiliation{Department of Physical Sciences,
 Indian Institute of Science Education and Research Kolkata, Mohanpur, West Bengal,741246, INDIA.}
\author{Partha Mitra}
 \email{pmitra@iiserkol.ac.in}
\affiliation{Department of Physical Sciences,
 Indian Institute of Science Education and Research Kolkata, Mohanpur, West Bengal,741246, INDIA.}

\date{\today}

\begin{abstract}
Harmonic Hall measurements were performed on a series of ferromagnetic metal/heavy metal (FM/HM) bilayers consisting of Permalloy (Py) as the FM and $\beta$-Tungsten (W) as the HM, and the efficiencies of the two orthogonal components of the spin-orbit torque (SOT) were extracted. Two sets of Hall bar-shaped devices, differing in the aspect ratio of the voltage pickup line width and the current channel width, were studied. Within each set, the resistivity of the W layer was systematically varied over a wide range (approximately 150–1000 $\mu\Omega\cdot$cm). To account for geometry-induced variations in current distribution, numerical simulations were performed, and a correction protocol was developed to normalize the torque efficiencies obtained from the conventional analysis. After applying the correction, the Slonczewski-like (anti-damping, in-plane) torque efficiency exhibited a consistent dependence on W resistivity across both device sets. In contrast, the field-like (out-of-plane) torque efficiency remained largely independent of W resistivity, reinforcing its interfacial character.

\end{abstract}

\maketitle

A bilayer comprising a ferromagnetic metal (FM) and a heavy metal (HM) serves as a simple yet powerful platform for experimentally validating key concepts in spin-orbitronics, most notably, inducing magnetization dynamics using electric fields. HM broadly refers to conductors with strong spin-orbit coupling (SOC), such as high atomic number elemental metals, topological insulators, and certain semiconductors. SOC is a crucial material property that enables the interconversion between charge and spin currents via the Spin Hall Effect (SHE) and its reciprocal, the Inverse Spin Hall Effect (ISHE) \cite{RevModPhys871213,jungwirth2012spin,hoffmann2013spin}. Presently two popular experimental routes  are pursued to study spintronic behavior of FM/HM bilayers. In ferromagnetic resonance (FMR) experiments \cite{PhysRevLett88117601,mizukami2002effect,saitoh2006conversion,mosendz2010quantifying}, the FM layer acts as a source of spin current through spin pumping from a precessing magnetization driven by an applied radio-frequency field. This spin current is injected into the adjacent HM layer and is converted into a measurable transverse voltage in the HM due to the ISHE. Conversely, in harmonic Hall (HH) measurements\cite{PhysRevB89144425,PhysRevB90224427}, an a.c. excitation flowing through the HM layer generates an oscillating spin current via the SHE which is injected into the adjacent FM layer. The component of the spin angular momentum current that is transverse to the magnetization is absorbed by the FM, exerting a torque known as Spin-Orbit Torque (SOT)\cite{PhysRevB78212405, PhysRevB79094422, PhysRevB80094424, PhysRevB86014416, Chernyshov2009, Miron2010, Miron2011, PhysRevLett101036601, PhysRevLett106036601,10106314863407}. In addition to bulk SHE-induced torques, SOT can also arise from interface effects due to the Rashba-Edelstein Effect (REE)\cite{EDELSTEIN1990233, PhysRevLett92256601}. In this mechanism, the in-plane charge current along the FM/HM interface generates spin accumulation at the interface due to Rashba-type SOC which exerts a torque on the FM layer, contributing further to the total SOT\cite{doi101098rsta20100336, Miron2011,10106314864399,10106314914897,PhysRevB91144412}. The interplay between the oscillating magnetization and the a.c. excitation in the FM gives rise to a characteristic second harmonic Hall voltage in the bilayer structure\cite{PhysRevB90224427}. Tuning the magnitude of SOT through careful material selection and device geometry optimization, is a key goal in advancing electric field–induced magnetization dynamics and ultimately to achieve magnetic field-free magnetization switching, which is essential for next-generation low-power spintronic devices.

The microscopic mechanisms responsible for SHE in HMs involve the spin-dependent preferential deflection of charge carriers in the transverse direction. These mechanisms are analogous to those responsible for the anomalous Hall effect (AHE) in FMs \cite{nagaosa2010anomalous}, a phenomenon that has been the subject of intensive research over several decades. that lead to the broad classification of the underlying mechanisms into two categories. \textit{Extrinsic} mechanisms arise from relativistic scattering of carriers by local electric fields at defect or impurity sites, whereas \textit{intrinsic} mechanisms originate purely from the band structure of the material and can exist even in the absence of defects or impurities. However, a similarly rigorous experimental investigation to conclusively establish the dominant mechanisms in SHE is still lacking.

A quantitative measure of SOC strength and hence the efficiency of spin-to-charge current interconversion, is given by the spin Hall angle ($\theta_{\mathrm{SH}}$), defined as the ratio of the generated spin current to the applied charge current. The sign of $\theta_{\mathrm{SH}}$ determine the spin polarization direction relative to the charge current direction. According to accepted conventions, $\theta_{\mathrm{SH}}$ is positive for platinum (Pt) and negative for tungsten (W). It has been reported \cite{pai2012spin} that the $\alpha$-phase W, a low-resistivity crystalline form, exhibits a much lower $\theta_{\mathrm{SH}}$ compared to the $\beta$-phase W, a high-resistivity metastable phase. This observation underscores the significant role of impurity scattering in spin current generation via the extrinsic mechanism. In contrast, the REE induced spin accumulation represents a special case of an intrinsic mechanism, as it arises from the broken inversion symmetry at the ferromagnet/heavy metal (FM/HM) interface and, in principle, should not depend on the resistivity of the heavy metal layer. Systematically varying the resistivity of the HM layer—while keeping the FM layer constant—and examining the resulting changes in spin–orbit torque (SOT) efficiency can provide valuable insights into the dominant spin current generation mechanisms in heavy metals. Notably, the tunability of resistivity over a wide range in the $\beta$-W phase \cite{hao2015beta} via simple growth condition adjustments offers a unique opportunity to explore these fundamental questions.

\begin{figure}[ht]
\includegraphics[width=.45\textwidth]{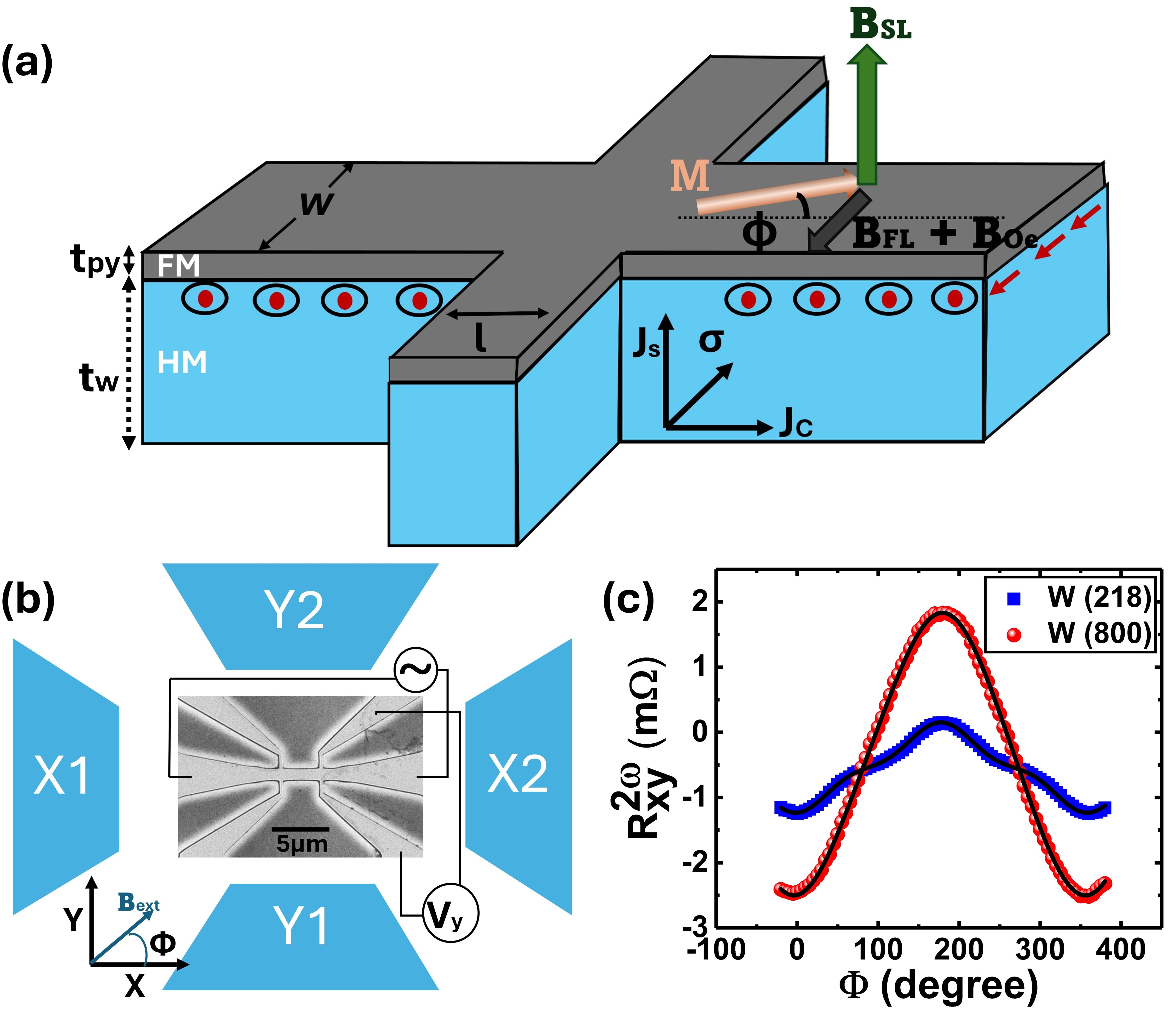}
\caption{ Schematic diagram of (a) SOT in FM/HM bilayer and (b) HH measurement setup overlaid with an SEM image of a typical device (top view). (c) Variation of second harmonic Hall resistance $R_{xy}^{2\omega}=V_y^{2\omega}/I_0$ for two devices  at $B_{ext}=0.1T$.}
\label{schematics}
\end{figure}

In earlier work, we reported a detailed FMR study on a series of Py/W bilayers, where the resistivity of the W layer ($\rho_W$) was systematically varied over a range of $100$–$1400~\mu\Omega \cdot \mathrm{cm}$, while keeping the Py layer constant \cite{Aon2024}. The measurements revealed that the effective Gilbert damping parameter, $\alpha_{\mathrm{eff}}$, increases monotonically with increasing $\rho_W$, indicative of enhanced spin current absorption in the W layer. This behavior suggests an increase in spin mixing conductance and/or spin Hall conductivity with resistivity, consistent with expectations if extrinsic mechanisms dominate. In the current work, we present a complementary investigation using harmonic Hall measurements \cite{PhysRevB90224427} (Fig.~\ref{schematics}) to extract the spin torque efficiency, which is directly related to $\theta_{\mathrm{SH}}$, although it remains influenced by the spin transparency of the FM/HM interface \cite{PhysRevLett116126601}. We have used three distinct device structures identified as Set 1, 2 and 3, described in Table ~\ref{SetParameters}. 

\begin{table}[hb]
\centering
\caption{Summary of device parameters for Py/W bilayer sets.}
\label{SetParameters}
\begin{tabular}{| c | c | c | c |}
\hline
\textbf{Parameter} & \textbf{Set 1} & \textbf{Set 2} & \textbf{Set 3} \\
\hline
$w\ (\mu m) \times L\ (\mu m)$ & $28\ \times 300$ & $2\ \times 20$ & $2\ \times 20$ \\
\hline
$l (\mu m)$ & $18$ & $0.5$ & $0.3$, $0.6$, $1$, \\
 & & & $1.6$, $2.2$, $2.6$ \\
\hline
$t_{\text{Py}}$ ( nm) & $10$ & $6$ & $6$ \\
\hline
$t_W$ ( nm) & $15$ & $15$  & $15$ \\
\hline
$\rho_{Py}(\mu\Omega\cdot\text{cm})$ & 76 & 90 & 90 \\
\hline
$\rho_W(\mu\Omega\cdot\text{cm})$ & 140, 350, & 218, 290, 454, & $\sim$450 \\
 & 520, 1025 & 800, 980 & \\
\hline
\end{tabular}
\label{tab:device_summary}
\end{table}

When a current flows along the FM/HM bilayer, the transverse voltage response primarily arises from the current flowing through the FM layer, due to the combined effects of the planar Hall effect (PHE) and the anomalous Hall effect (AHE). The first harmonic voltage response is determined by the equilibrium magnetization state, which is controlled by the external magnetic field. For FMs with strong in-plane anisotropy, such as Py, the PHE is the dominant contribution, described by $R_{xy} ^ {\omega}=R_{PHE}sin(2\Phi)$, where $\Phi$ is the angle between the magnetization and current flow direction ( See Fig. \ref{schematics}). In contrast, the second harmonic voltage arises from a perturbative tilting of the magnetization about its equilibrium position, induced by spin–orbit torque (SOT) \cite{PhysRevB100214438,PhysRevB89144425}. This torque originates from the spin current diffusing from the HM or spin accumulation at the HM/FM interface, along with additional contributions from heating effects \cite{PhysRevB90224427}. The out-of-plane tilt contributes to the AHE component, while the in-plane deviation modifies the PHE component. For in-plane anisotropic ferromagnets like Py, the out-of-plane and in-plane tilts correspond to the field-like (FL) and antidamping or Slonczewski-like (SL) torques, respectively, as described by the following equation\cite{aoki2023gigantic}\cite{PhysRevB99195103}\cite{AHN202312}:

\begin{equation}
\begin{split}
    R_{xy}^{2\omega}(\Phi)=[R_{AHE}(\frac{B_{SL}}{B_{ext}+B_{eff}^k})+\alpha B_{ext} + R_{\nabla T}]cos(\Phi)\\
    +2R_{PHE}(\frac{B_{FL}+B_{Oe}}{B_{ext}})cos(2\Phi)cos(\Phi)
    \end{split}
    \label{eq:Rxy2omega}
\end{equation}

In a typical  HH experiment, for an a.c. excitation of fixed amplitude, the magnetization is rotated fully in the plane of the device relative to the current direction by applying an external magnetic field of fixed magnitude ($B_{ext}$). During this in-plane rotation, SOT varies systematically, which manifests as a characteristic angular dependence in the second harmonic voltage. By performing such angular scans over a range of $B_{ext}$ magnitudes, the SOT effective fields—field-like ($B_{FL}$) and Slonczewski-like or antidamping-like ($B_{SL}$)—can be extracted using Eq.~\ref{eq:Rxy2omega}. To facilitate quantitative comparison between different devices, these SOT fields are normalized with respect to the saturation magnetization ($M_s$) of the ferromagnetic layer and the current density in the heavy metal layer ($J_{HM}$), yielding the spin torque efficiencies defined as\cite{PhysRevB87020402}\cite{PhysRevB92064426}:

\begin{equation}
   \xi_{SL(FL)} =\frac{2e}{\hbar}\frac{B_{SL(FL)}M_st_{FM}}{J_{HM}}
    \label{eq:zeta}
\end{equation}

The amplitude of the a.c. excitation current($I_o$) is calculated using a parallel resistor model for the FM and HM layers, ensuring that the current density in the tungsten layer ($J_W$) remains the same across all devices within a given set. For Set 1, $J_W = 5.6 \times 10^{10} ,\mathrm{A/m^2}$, while for Set 2, $J_W = 7.9 \times 10^{10} ,\mathrm{A/m^2}$. This approach was adopted to account for the systematic increase in $\rho_W$ within each set, while keeping the resistivity of Py ($\rho_{Py}$) and all other geometric parameters constant. As a result, the current flowing through the HM layer decreases significantly relative to that through the FM layer with increasing $\rho_W$. Maintaining a constant $J_W$ within a device set facilitates direct comparison of spin current generation across different W layers and ensures the Oersted field acting on the FM layer remains constant. The anomalous Hall resistance ($R_{AHE}$) and planar Hall resistance ($R_{PHE}$) were obtained from the first harmonic voltage response measured under identical current excitation conditions. Our analysis reveals that the magnitude of SL torque efficiency ($\xi_{SL}$) increases systematically with the resistivity of the W layer (Fig.~\ref{unscaled zeta sl}) for both sets of devices, although the values for Set 1 were consistently lower than those of Set 2. This observation supports the idea that enhanced electron scattering in the high-resistivity $\beta$-W phase leads to increased spin current generation, which manifests as an increase in the SL component of the SOT. In contrast, FL torque efficiency ($\xi_{FL}$) showed no pronounced dependence on $\rho_W$ (Fig.~\ref{unscaled zeta fl}), and the values for Set 1 were consistently higher in magnitude than those for Set 2. This behavior can be explained by recognizing that the FL torque predominantly arises from the Rashba–Edelstein effect (REE)\cite{haney2013current,10106315027855}, which is governed mainly by the FM/HM interface properties—these remain broadly constant within a given device set. Finally, note that the negative values of $\xi$ arise from the convention that the spin Hall angle ($\theta_{SHE}$) of tungsten is taken to be negative. The systematic shift in the $\xi$ values between Set 1 and Set 2 with different geometric structures suggests an influence of device dimensions on the absolute values of the measured SOT efficiencies. This effect has been previously reported \cite{10106315037391} and is attributed to current density  "thinning" in the region between the voltage pickup lines, which leads to a reduction in the effective current density contributing to the spin–orbit torque.

\begin{figure}[h]
    \begin{subfigure}[t]{0.485\columnwidth} 
        \centering
        \includegraphics[width=\linewidth]{image/SLefficiencyOfBothSet.jpg}
         \caption{}
        \label{unscaled zeta sl}
    \end{subfigure}
    \hfill
    \begin{subfigure}[t]{0.5\columnwidth}
        \centering
        \includegraphics[width=\linewidth]{image/FLefficiencyOfBothSet.jpg}
        \caption{}
        \label{unscaled zeta fl}
    \end{subfigure}
    
    \caption{Variation of spin–orbit torque (SOT) efficiencies with resistivity for devices in Set 1 (aspect ratio 0.26, blue spheres) and Set 2 (aspect ratio 0.71, green squares): (a) Damping-like torque efficiency $\xi_{SL}$ and (b) Field-like torque efficiency $\xi_{FL}$.}
    \label{unscaled zeta vs rho}
\end{figure}

\begin{figure}[h]
    \begin{subfigure}[t]{0.49\columnwidth} 
        \centering
        \includegraphics[width=\linewidth]{image/ZetaVslbyW.jpg}
        \caption{}
        \label{fig:subim1}
    \end{subfigure}
    \hfill
    \begin{subfigure}[t]{0.495\columnwidth}
        \centering
        \includegraphics[width=\linewidth]{image/ZetaFLvsLbyW.jpg}
        \caption{}
        \label{fig:subim2}
    \end{subfigure}
    \begin{subfigure}[t]{\columnwidth} 
        \centering
        \includegraphics[width=\linewidth]{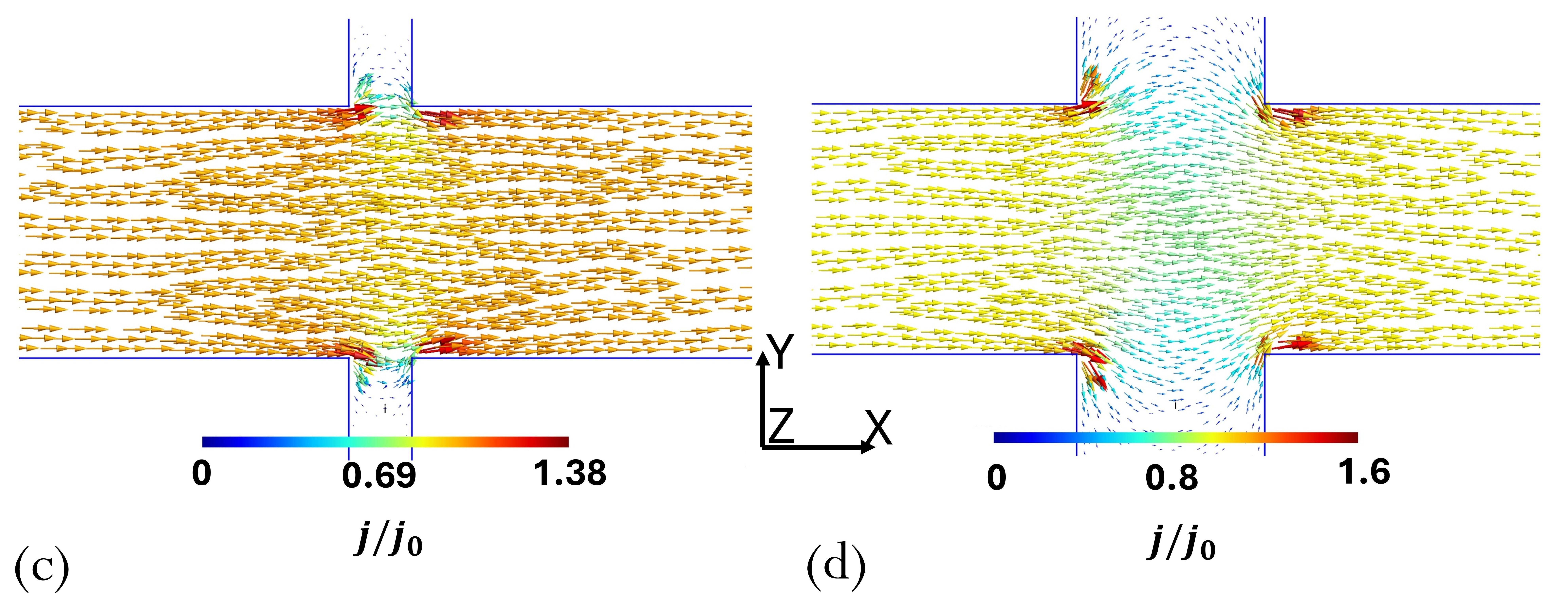}
        \label{fig:subim1}
    \end{subfigure}
    
   \caption{(a) SL torque efficiency $\xi_{SL}$ and (b) FL torque efficiency $\xi_{FL}$ as a function of aspect ratio ($l/w$). The red curves represent the predicted $\xi_{SL/FL}$ values obtained by scaling the value at the lowest aspect ratio using the simulated current density reduction (Eq.~\ref{eq:scaled zeta}) for each corresponding device. The blue curve in (b) represents the simulated $\xi_{SL}$ including corrections due to the Oersted field. 
(c) and (d) Simulated current distribution in Hall bars with aspect ratios of 0.25 and 0.75, respectively, using the GETDP solver.
} 
    \label{fig:vpl-variation}
\end{figure}

To investigate the origin of the overall shift in SOT efficiency values observed between Set 1 and Set 2 with differing geometric aspect ratios ($l/w$), we fabricated a third series of Hall bar devices, denoted as Set 3, where the length of the voltage pickup leads ($l$) was systematically varied from $0.25\mu m$ to $2.5\mu m$, while keeping the channel width ($w$) fixed at $2\mu m$. This effectively varied the aspect ratio $l/w$ across a range of 0.125 to 1.125. The effective fields and corresponding SOT efficiencies were extracted using the same procedure described previously. The results revealed a pronounced decrease in both $\xi_{SL}$ (Fig. \ref{fig:vpl-variation}(a)) and $\xi_{FL}$(Fig. \ref{fig:vpl-variation}(b))  with increasing $l/w$ . To gain further insight on the effect of voltage lead aspect ratio, we conducted finite-element simulations using GMSH software to explore the current distribution in the current channel in the presence of voltage leads. A 3D model of a Hall bar was created and a finite-element mesh, consisting of small elements representing the geometry of the Hall bar, was generated within the same environment. Using the GETDP solver, the Laplace equation ($\nabla^2 V=0$) was solved under the boundary condition of an applied potential difference across the ends of the current channel. This simulation provided the electric potential distribution across all mesh elements. From the potential, the electric field was computed using $\vec{E}=-\nabla \vec V$, and subsequently, the current density was obtained using $\vec{j}=\sigma\vec{E}$, where $\sigma$ is the electrical conductivity. 
The resultant current distribution was visualized using arrows, where the areal density and color of arrows represent the magnitude of the local current density as shown in Fig \ref{fig:vpl-variation} (c) and (d) for structures of aspect ratios of 0.25 and 0.75 respectively. The figures clearly indicates a non-uniform current distribution along the channel near the crossed region between voltage pick up leads the current density was found to be reduced compared to the region far away at the ends.  From visual inspection it is apparent that for lower aspect ratio  Hall bars (Fig. \ref{fig:vpl-variation}(c)) , this reduction of  current density is less compared to the case of relatively larger aspect ratios(Fig. \ref{fig:vpl-variation}(d)). 
The simulations were carried out over a large parameter space of the aspect ratio and for each structure the reduction of current density within the region of voltage leads is quantified by extracting the ratio of the  root mean square (RMS) value of the current in the cross region of the Hall bar $\sqrt{<{(j^2(\vec{r}))}>}$  to the current value far from the cross region $j_o$. 

The procedure for calculating the SOT efficiencies  $\xi_{SL/FL}$ (Fig.\ref{unscaled zeta vs rho}) using Eq.\ref{eq:zeta}, where the current density value used is the ideal value ($j_o$) which holds only in the region far from the voltage lines. But since the current density responsible for the observed $2\omega$ response  is the one that exists between the voltage lines, a correction factor is necessary to account for the discrepancy to estimate the actual value of $\xi$.  
To establish connection between experimental values from Set 3 and the simulation results, we choose the $\xi_{SL}$ value of the lowest aspect ratio device in Set 3 ($l/w=0.15$) as a reference, where the simulations showed that $\sqrt{<{(j^2(\vec{r}))}>}/j_o= 0.94$, 
and predicted the values of $\xi_{SL/FL}$ for the rest of the devices in the set using the simulated values of the current density ratio of the corresponding structures. The results of the simulation are shown as thick red lines in Fig \ref{fig:vpl-variation} (a) and (b). We observe that for $\xi_{SL}$ , the experimental data and the simulated values are in close agreement, confirming that the bulk current density indeed controls the in-plane SOT component. However, for $\xi_{FL}$ we find that the simulated values are consistently lower in magnitude compared to the experimental values. Eq.\ref{eq:Rxy2omega} shows that $B_{FL}$ and the Oersted field   $B_{Oe}$ both has same symmetry and generates out-of-plane torque and both are directly proportional to current density in the $W$ layer. Hence, while extracting the FL efficiency, the corrected value of Oersted field needs to be considered and the results are shown in thick blue lines in Fig.\ref{fig:vpl-variation}(b) and is in reasonably good agreement with the experimental data.

The correlation between current density "thinning" within the voltage leads and reduction in the magnitude of $\xi_{SL/FL}$ with increasing $l/w$ ratio is applied to the data shown in Fig \ref{unscaled zeta vs rho} which shows an offset in the two sets of devices. From the simulations, the current density reduction ratio $\sqrt{<{(j^2(\vec{r}))}>}/j_o$ is obtained to be 0.90 and 0.79 for $l/w$ of 0.26 (Set 2) and 0.71 (Set 1) respectively. We define a scaled SOT efficiency as 
\begin{equation}
   \xi_{SL(FL)}^\prime =\frac{\xi_{SL(FL)}}{(\sqrt{<{(j^2(\vec{r}))}>}/j_o)}
    \label{eq:scaled zeta}
\end{equation}
 which signifies the ideal value of SOT efficiency arising form the ideal current density $j_o$. The plot of the $\xi_{SL}^\prime$ for both Set 1 and Set 2 devices as a function of $\rho_W$  on a logarithmic scale seems to follow the same underlying trend (Fig.\ref{fig:scaled zeta sl}), barring one device with the lowest resistivity ($\rho_W=140\mu\Omega.cm$). The magnitude of $\xi_{SL}^\prime$ exhibits a 'weak' dependence on $\rho_W$, showing approximately a 60\% enhancement for one order of magnitude variation of $\rho_W$. 
 
 A model for  SL torque arising from SHE in the HM layer was developed in a previous report \cite{PhysRevLett116126601} where $\xi_{SL}$ was expressed  in terms of the bulk spin Hall angle $\theta_{SH}=(\sigma_{SH}/\sigma)$, correction factors for finite thickness of the HM (that tends to unity for thickness much greater spin diffusion length $\lambda_s$) and  transparency parameter of the FM/HM interface characterized by spin mixing conductance and resistivity. In this report \cite{PhysRevLett116126601}, the dependence of $\xi_{SL}$ on a set of Pt films with varying resistivity ($20-100 \mu\Omega .cm$)  controlled by film thicknesses ($3-15nm$) was explained using the model of intrinsic mechanism of SHE where the $\sigma_{SH}$ is assumed to be independent of resistivity. In a subsequent report\cite{10106315027855}, the variation of  $\xi_{SL}$  was measured on a series of W films of varying thickness $(1-6nm)$. The effective spin Hall angle (that includes the contribution of interface transparency) assumed to be same for the entire set was reported to be $\theta^{eff}_{SH}\sim0.6$. In contrast, we stabilize the $\beta-$phase W at a much larger nominal thickness ($\sim15 nm$), keeping thickness constant while varying resistivity solely by adjusting the Ar sputtering pressure. This ensures that  the films are homogeneous with well defined interface, resistivity variation originates primarily from bulk scattering and  the W thickness significantly exceeds the spin-diffusion length so that spin accumulation reaches its saturation regime. Our attempt to apply the model of intrinsic mechanism to our data set did not yield meaningful fitting parameters , indicating incompatibility of the model. Hence, we modified the model  assuming $\theta_{SH}$ to be constant and independent of $\rho_W$ \cite{haney2013current,PhysRevB87144411,PhysRevLett116126601}:

\begin{equation}
\xi_{SL}^\prime = \frac{\theta_{SH} (2\lambda_s G_r)\rho_W}{1 + (2\lambda_s G_r) \rho_W}
\label{eq:zetaSlvsthetaSH}
\end{equation}

Here, $G_r$ is the real part of the spin mixing conductance $G^{\uparrow\downarrow} = G_r + iG_i$, with $G_r \gg G_i$. As shown in  Figure \ref{fig:scaled zeta fit}, a numerical  fit using Eq. \ref{eq:zetaSlvsthetaSH},  barring the data point at the lowest resistivity, which is corresponds to a mixed $\alpha+\beta$ phase W, appears to capture the variation for the pure $\beta-W$ samples. The numerical fit  results in $\theta_{SH} = 0.6 \pm 0.1$ and $G_r \approx 10^{14},\Omega^{-1}\mathrm{m}^{-2}$, assuming a spin diffusion length $\lambda_s \sim 3,\text{nm}$. Previous FMR measurements on  W-Py bilayers \cite{cao2019measurement} reports $\lambda_s <2nm$ and $G_r \sim 2.8 \times10^{14}\Omega^{-1}m^{-2}$ irrespective of the amounts of $\alpha-$ or $\beta-$ phase present in the W films, which is in good agreement with our results. 
Further, to account for combination of both categories of mechanisms , we considered $\theta_{SH}=a + b\rho_W$. However,  the fitting results are poor and doesnot reproduce the data as accurately as the pure skew scattering case.

The FL torque efficiency $\xi_{FL}$, as shown in Fig. \ref{unscaled zeta fl},  does not show any pronounced dependence on $\rho_W$ in both device sets, which is also consistent with the earlier report\cite{10106315027855}. For metallic systems, it is generally accepted that the FL torque is dominated the Rashba–Edelstein effect (REE), and is given by \cite{PhysRevB87020402}\cite{haney2013current}:

\begin{equation}
\xi_{FL}^\prime = \frac{\alpha_R P t_{FM}}{2\mu_B}
\label{eq:zetaREE}
\end{equation}

Here, $\alpha_R$ is the Rashba coefficient, $\mu_B$ is the Bohr magneton, $P$ is the spin polarization of the ferromagnet (FM), and $t_{FM}$ is the thickness of the FM layer. Since the FM layer thicknesses in Set 1 and Set 2 differ, we further normalized the FL efficiency by $t_{FM}$, resulting in the quantity $\xi^\prime_{FL}/t_{FM}$. As shown in the inset of Fig. \ref{fig:scaled zeta sl}, this normalization removes the offset observed in the unscaled $\xi_{FL}$ values (Fig. \ref{unscaled zeta fl}), and both data sets collapse onto a single underlying trend, as expected.

\begin{figure}[h!]
    \begin{subfigure}[t]{0.95\columnwidth} 
        \centering
        \includegraphics[width=\linewidth]{image/ScaledZeta.jpg}
         \caption{}
        \label{fig:scaled zeta sl}
    \end{subfigure}
    \hfill
    \begin{subfigure}[t]{0.95\columnwidth}
        \centering
        \includegraphics[width=\linewidth]{image/ZetaVsRho2.jpg}
        \caption{}
        \label{fig:scaled zeta fit}
    \end{subfigure}
    
    \caption{ (a) Variation of scaled spin–orbit torque (SOT) efficiency $\xi^\prime_{SL}$ (Eq.\ref{eq:scaled zeta}) for Sets 1 and 2 as a function of $\rho_W$ (in log scale), showing a consistent trend across devices, except for the one with the lowest $\rho_W$. Inset: Scaled field-like (FL) torque efficiency $\xi^\prime_{FL}$ normalized by the thickness of the FM layer.  (b) Fit of the magnitude of $\xi^\prime_{SL}$ versus $\rho_W$ using Eq.~\ref{eq:zetaSlvsthetaSH}, excluding the device with the lowest $\rho_W$.}
    \label{fig:scaled zeta vs rho}
\end{figure}

To conclude, our systematic investigation of Harmonic Hall analysis on a large set Py/W bilayer devices confirms that SL torque efficiency varies with tungsten resistivity validating its origin in Spin Hall effect of tungsten. Conversely, field-like torque efficiency remains invariant with resistivity changes, consistent with its interfacial origin and independence from bulk scattering mechanisms.
Our work further confirms that, geometric factors—particularly aspect ratio and voltage pickup line width variations significantly influences quantitative estimation of  spin-orbit torque efficiency. Finite-element simulations confirm that these geometric parameters alter current distribution patterns, directly impacting the calculation and hence introducing systematic deviation in  torque efficiency values. These findings establish that accurate spin-orbit torque characterization requires simultaneous consideration of both material properties and device geometry, providing a comprehensive framework for optimizing spintronic device performance through coordinated material resistivity engineering and structural design optimization.

See the Supplementary Information for further details on XRD characterization of thin films ( Note 1), device fabrication details (Note 2),  measurement procedures and data analysis (Note 3), simulation details (Note 4), and tabulated data corresponding to the key plots presented in the manuscript.

The authors thank IISER Kolkata, an autonomous research and teaching institution funded
by the MoE, Government of India for providing the financial support and infrastructure. The authors also thank CSIR and UGC for providing fellowship.

\bibliography{aipsamp}

@PREAMBLE{
 "\providecommand{\noopsort}[1]{}" 
 # "\providecommand{\singleletter}[1]{#1}%" 
}

@article{RevModPhys871213,
  title = {Spin Hall effects},
  author = {Sinova, Jairo and Valenzuela, Sergio O. and Wunderlich, J. and Back, C. H. and Jungwirth, T.},
  journal = {Rev. Mod. Phys.},
  volume = {87},
  issue = {4},
  pages = {1213--1260},
  numpages = {47},
  year = {2015},
  month = {Oct},
  publisher = {American Physical Society},
  doi = {10.1103/RevModPhys.87.1213},
  url = {https://link.aps.org/doi/10.1103/RevModPhys.87.1213}
}

@article{jungwirth2012spin,
  title={Spin Hall effect devices},
  author={Jungwirth, Tomas and Wunderlich, J{\"o}rg and Olejn{\'\i}k, Kamil},
  journal={Nature materials},
  volume={11},
  number={5},
  pages={382--390},
  year={2012},
  publisher={Nature Publishing Group UK London}
}

@article{hoffmann2013spin,
  title={Spin Hall effects in metals},
  author={Hoffmann, Axel},
  journal={IEEE transactions on magnetics},
  volume={49},
  number={10},
  pages={5172--5193},
  year={2013},
  publisher={IEEE}
}

@article{PhysRevLett88117601,
  title = {Enhanced Gilbert Damping in Thin Ferromagnetic Films},
  author = {Tserkovnyak, Yaroslav and Brataas, Arne and Bauer, Gerrit E. W.},
  journal = {Phys. Rev. Lett.},
  volume = {88},
  issue = {11},
  pages = {117601},
  numpages = {4},
  year = {2002},
  month = {Feb},
  publisher = {American Physical Society},
  doi = {10.1103/PhysRevLett.88.117601},
  url = {https://link.aps.org/doi/10.1103/PhysRevLett.88.117601}
}

@article{mizukami2002effect,
  title={Effect of spin diffusion on Gilbert damping for a very thin permalloy layer in Cu/permalloy/Cu/Pt films},
  author={Mizukami, Shigemi and Ando, Yasuo and Miyazaki, Terunobu},
  journal={Physical Review B},
  volume={66},
  number={10},
  pages={104413},
  year={2002},
  publisher={APS}
}

@article{saitoh2006conversion,
  title={Conversion of spin current into charge current at room temperature: Inverse spin-Hall effect},
  author={Saitoh, E and Ueda, M and Miyajima, H and Tatara, G},
  journal={Applied physics letters},
  volume={88},
  number={18},
  year={2006},
  publisher={AIP Publishing}
}

@article{mosendz2010quantifying,
  title={Quantifying spin Hall angles from spin pumping: Experiments and theory},
  author={Mosendz, O and Pearson, JE and Fradin, FY and Bauer, GEW and Bader, SD and Hoffmann, A},
  journal={Physical review letters},
  volume={104},
  number={4},
  pages={046601},
  year={2010},
  publisher={APS}
}

@article{PhysRevB89144425,
  title = {Quantitative characterization of the spin-orbit torque using harmonic Hall voltage measurements},
  author = {Hayashi, Masamitsu and Kim, Junyeon and Yamanouchi, Michihiko and Ohno, Hideo},
  journal = {Phys. Rev. B},
  volume = {89},
  issue = {14},
  pages = {144425},
  numpages = {15},
  year = {2014},
  month = {Apr},
  publisher = {American Physical Society},
  doi = {10.1103/PhysRevB.89.144425},
  url = {https://link.aps.org/doi/10.1103/PhysRevB.89.144425}
}

@article{PhysRevB90224427,
  title = {Interplay of spin-orbit torque and thermoelectric effects in ferromagnet/normal-metal bilayers},
  author = {Avci, Can Onur and Garello, Kevin and Gabureac, Mihai and Ghosh, Abhijit and Fuhrer, Andreas and Alvarado, Santos F. and Gambardella, Pietro},
  journal = {Phys. Rev. B},
  volume = {90},
  issue = {22},
  pages = {224427},
  numpages = {11},
  year = {2014},
  month = {Dec},
  publisher = {American Physical Society},
  doi = {10.1103/PhysRevB.90.224427},
  url = {https://link.aps.org/doi/10.1103/PhysRevB.90.224427}
}

@article{PhysRevB78212405,
  title = {Theory of nonequilibrium intrinsic spin torque in a single nanomagnet},
  author = {Manchon, A. and Zhang, S.},
  journal = {Phys. Rev. B},
  volume = {78},
  issue = {21},
  pages = {212405},
  numpages = {4},
  year = {2008},
  month = {Dec},
  publisher = {American Physical Society},
  doi = {10.1103/PhysRevB.78.212405},
  url = {https://link.aps.org/doi/10.1103/PhysRevB.78.212405}
}

@article{PhysRevB79094422,
  title = {Theory of spin torque due to spin-orbit coupling},
  author = {Manchon, A. and Zhang, S.},
  journal = {Phys. Rev. B},
  volume = {79},
  issue = {9},
  pages = {094422},
  numpages = {9},
  year = {2009},
  month = {Mar},
  publisher = {American Physical Society},
  doi = {10.1103/PhysRevB.79.094422},
  url = {https://link.aps.org/doi/10.1103/PhysRevB.79.094422}
}

@article{PhysRevB80094424,
  title = {Spin-orbit coupling mediated spin torque in a single ferromagnetic layer},
  author = {Matos-Abiague, A. and Rodr\'{\i}guez-Su\'arez, R. L.},
  journal = {Phys. Rev. B},
  volume = {80},
  issue = {9},
  pages = {094424},
  numpages = {6},
  year = {2009},
  month = {Sep},
  publisher = {American Physical Society},
  doi = {10.1103/PhysRevB.80.094424},
  url = {https://link.aps.org/doi/10.1103/PhysRevB.80.094424}
}

@article{PhysRevB86014416,
  title = {Quantum kinetic theory of current-induced torques in Rashba ferromagnets},
  author = {Pesin, D. A. and MacDonald, A. H.},
  journal = {Phys. Rev. B},
  volume = {86},
  issue = {1},
  pages = {014416},
  numpages = {5},
  year = {2012},
  month = {Jul},
  publisher = {American Physical Society},
  doi = {10.1103/PhysRevB.86.014416},
  url = {https://link.aps.org/doi/10.1103/PhysRevB.86.014416}
}

@article{Chernyshov2009,
   title        = {Evidence for reversible control of magnetization in a ferromagnetic material by means of spin-orbit magnetic field},
   author       = {Chernyshov, A. and Overby, M. and  Liu, X. and Furdyna, J.K. and Lyanda-Geller, Y. and Leonid, P.R.},
   journal      = {Nature Physics},
   year         = {2009}, 
   volume       = {5}, 
   pages        = {9},
   publisher = {Nature},
  doi = {https://doi.org/10.1038/nphys1362},
  url = {https://www.nature.com/articles/nphys1362}
}

@article{Miron2010,
   title        = {Current-driven spin torque induced by the Rashba effect in a ferromagnetic metal layer.},
   author       = {Miron, I.M. and Gaudin, G. and Auffret, S. and Rodmac, B. and Schuhl, A. and Pizzini, S. and Vogel, J. and Gambardella, P.},
   journal      = {Nature materials},
   year         = {2010}, 
   volume       = {9}, 
   pages        = {3},
   publisher = {Nature},
  doi = {https://doi.org/10.1038/nmat2613},
  url = {https://www.nature.com/articles/nmat2613}
}

@article{Miron2011,
   title        = {Perpendicular switching of a single ferromagnetic layer induced by in-plane current injection},
   author       = {Miron, I.M. and Garello, K. and Gaudin, G. and Zermatten, P.J. and Costache, M.V. and Auffret, S. and Bandiera, S. and  Rodmac, B. and Schuhl, A. and Gambardella, P.},
   journal      = {Nature},
   year         = {2011}, 
   volume       = {476}, 
   pages        = {7359},
   publisher = {Nature},
  doi = {https://doi.org/10.1038/nature10309},
  url = {https://www.nature.com/articles/nature10309%7D}
}

@article{PhysRevLett101036601,
  title = {Electric Manipulation of Spin Relaxation Using the Spin Hall Effect},
  author = {Ando, K. and Takahashi, S. and Harii, K. and Sasage, K. and Ieda, J. and Maekawa, S. and Saitoh, E.},
  journal = {Phys. Rev. Lett.},
  volume = {101},
  issue = {3},
  pages = {036601},
  numpages = {4},
  year = {2008},
  month = {Jul},
  publisher = {American Physical Society},
  doi = {10.1103/PhysRevLett.101.036601},
  url = {https://link.aps.org/doi/10.1103/PhysRevLett.101.036601}
}

@article{PhysRevLett106036601,
  title = {Spin-Torque Ferromagnetic Resonance Induced by the Spin Hall Effect},
  author = {Liu, Luqiao and Moriyama, Takahiro and Ralph, D. C. and Buhrman, R. A.},
  journal = {Phys. Rev. Lett.},
  volume = {106},
  issue = {3},
  pages = {036601},
  numpages = {4},
  year = {2011},
  month = {Jan},
  publisher = {American Physical Society},
  doi = {10.1103/PhysRevLett.106.036601},
  url = {https://link.aps.org/doi/10.1103/PhysRevLett.106.036601}
}

@article{10106314863407,
    author = {Cubukcu, Murat and Boulle, Olivier and Drouard, Marc and Garello, Kevin and Onur Avci, Can and Mihai Miron, Ioan and Langer, Juergen and Ocker, Berthold and Gambardella, Pietro and Gaudin, Gilles},
    title = "{Spin-orbit torque magnetization switching of a three-terminal perpendicular magnetic tunnel junction}",
    journal = {Applied Physics Letters},
    volume = {104},
    number = {4},
    pages = {042406},
    year = {2014},
    month = {01},
    doi = {10.1063/1.4863407},
    url = {https://doi.org/10.1063/1.4863407},
}

@article{EDELSTEIN1990233,
title = {Spin polarization of conduction electrons induced by electric current in two-dimensional asymmetric electron systems},
author = {V.M. Edelstein},
journal = {Solid State Communications},
volume = {73},
number = {3},
pages = {233-235},
year = {1990},
issn = {0038-1098},
doi = {https://doi.org/10.1016/0038-1098(90)90963-C},
url = {https://www.sciencedirect.com/science/article/pii/003810989090963C},
}

@article{PhysRevLett92256601,
  title = {Experimental Separation of Rashba and Dresselhaus Spin Splittings in Semiconductor Quantum Wells},
  author = {Ganichev, S. D. and Bel'kov, V. V. and Golub, L. E. and Ivchenko, E. L. and Schneider, Petra and Giglberger, S. and Eroms, J. and De Boeck, J. and Borghs, G. and Wegscheider, W. and Weiss, D. and Prettl, W.},
  journal = {Phys. Rev. Lett.},
  volume = {92},
  issue = {25},
  pages = {256601},
  numpages = {4},
  year = {2004},
  month = {Jun},
  publisher = {American Physical Society},
  doi = {10.1103/PhysRevLett.92.256601},
  url = {https://link.aps.org/doi/10.1103/PhysRevLett.92.256601}
}

@article{doi101098rsta20100336,
author = {Gambardella, Pietro  and Miron, Ioan Mihai },
title = {Current-induced spin-orbit torques},
journal = {Philosophical Transactions of the Royal Society A: Mathematical, Physical and Engineering Sciences},
volume = {369},
number = {1948},
pages = {3175-3197},
year = {2011},
doi = {10.1098/rsta.2010.0336},

URL = {https://royalsocietypublishing.org/doi/abs/10.1098/rsta.2010.0336},
}

@article{10106314864399,
    author = {Skinner, T. D. and Wang, M. and Hindmarch, A. T. and Rushforth, A. W. and Irvine, A. C. and Heiss, D. and Kurebayashi, H. and Ferguson, A. J.},
    title = {Spin-orbit torque opposing the Oersted torque in ultrathin Co/Pt bilayers},
    journal = {Applied Physics Letters},
    volume = {104},
    number = {6},
    pages = {062401},
    year = {2014},
    month = {02},
    issn = {0003-6951},
    doi = {10.1063/1.4864399},
    url = {https://doi.org/10.1063/1.4864399},
}

@article{10106314914897,
    author = {Kawaguchi, M. and Moriyama, T. and Koyama, T. and Chiba, D. and Ono, T.},
    title = {Layer thickness dependence of current induced effective fields in ferromagnetic multilayers},
    journal = {Journal of Applied Physics},
    volume = {117},
    number = {17},
    pages = {17C730},
    year = {2015},
    month = {03},
    issn = {0021-8979},
    doi = {10.1063/1.4914897},
    url = {https://doi.org/10.1063/1.4914897},
}

@article{PhysRevB91144412,
  title = {Experimental demonstration of the coexistence of spin Hall and Rashba effects in $\ensuremath{\beta}\ensuremath{-}\text{tantalum/ferromagnet}$ bilayers},
  author = {Allen, Gary and Manipatruni, Sasikanth and Nikonov, Dmitri E. and Doczy, Mark and Young, Ian A.},
  journal = {Phys. Rev. B},
  volume = {91},
  issue = {14},
  pages = {144412},
  numpages = {9},
  year = {2015},
  month = {Apr},
  publisher = {American Physical Society},
  doi = {10.1103/PhysRevB.91.144412},
  url = {https://link.aps.org/doi/10.1103/PhysRevB.91.144412}
}

@article{nagaosa2010anomalous,
  title={Anomalous hall effect},
  author={Nagaosa, Naoto and Sinova, Jairo and Onoda, Shigeki and MacDonald, Allan H and Ong, Nai Phuan},
  journal={Reviews of modern physics},
  volume={82},
  number={2},
  pages={1539--1592},
  year={2010},
  publisher={APS}
}

@article{pai2012spin,
  title={Spin transfer torque devices utilizing the giant spin Hall effect of tungsten},
  author={Pai, Chi-Feng and Liu, Luqiao and Li, Y and Tseng, HW and Ralph, DC and Buhrman, RA},
  journal={Applied Physics Letters},
  volume={101},
  number={12},
  year={2012},
  publisher={AIP Publishing}
}

@article{hao2015beta,
  title={Beta ($\beta$) tungsten thin films: Structure, electron transport, and giant spin Hall effect},
  author={Hao, Qiang and Chen, Wenzhe and Xiao, Gang},
  journal={Applied Physics Letters},
  volume={106},
  number={18},
  year={2015},
  publisher={AIP Publishing}
}

@article{Aon2024,
doi = {10.1088/1402-4896/ad48c4},
url = {https://dx.doi.org/10.1088/1402-4896/ad48c4},
year = {2024},
month = {may},
publisher = {IOP Publishing},
volume = {99},
number = {6},
pages = {065546},
author = {Aon, Soumik and Pal, Sayani and Manna, Subhadip and Mitra, Chiranjib and Mitra, Partha},
title = {Modulating spin current induced effective damping in {$\beta$}–W/Py heterostructures by a systematic variation in resistivity of the sputtered deposited {$\beta$}–W films},
journal = {Physica Scripta},
}

@article{PhysRevLett116126601,
  title = {Spin Torque Study of the Spin Hall Conductivity and Spin Diffusion Length in Platinum Thin Films with Varying Resistivity},
  author = {Nguyen, Minh-Hai and Ralph, D. C. and Buhrman, R. A.},
  journal = {Phys. Rev. Lett.},
  volume = {116},
  issue = {12},
  pages = {126601},
  numpages = {6},
  year = {2016},
  month = {Mar},
  publisher = {American Physical Society},
  doi = {10.1103/PhysRevLett.116.126601},
  url = {https://link.aps.org/doi/10.1103/PhysRevLett.116.126601}
}

@article{PhysRevB100214438,
  title = {Elimination of thermoelectric artifacts in the harmonic Hall measurement of spin-orbit torque},
  author = {Park, Eun-Sang and Lee, Dong-Kyu and Min, Byoung-Chul and Lee, Kyung-Jin},
  journal = {Phys. Rev. B},
  volume = {100},
  issue = {21},
  pages = {214438},
  numpages = {11},
  year = {2019},
  month = {Dec},
  publisher = {American Physical Society},
  doi = {10.1103/PhysRevB.100.214438},
  url = {https://link.aps.org/doi/10.1103/PhysRevB.100.214438}
}

@article{aoki2023gigantic,
  title={Gigantic Anisotropy of Self-Induced Spin-Orbit Torque in Weyl Ferromagnet Co2MnGa},
  author={Aoki, Motomi and Yin, Yuefeng and Granville, Simon and Zhang, Yao and Medhekar, Nikhil V and Leiva, Livio and Ohshima, Ryo and Ando, Yuichiro and Shiraishi, Masashi},
  journal={Nano Letters},
  volume={23},
  number={15},
  pages={6951--6957},
  year={2023},
  publisher={ACS Publications}
}

@article{PhysRevB99195103,
  title = {Spin-orbit torque and Nernst effect in $\mathrm{Bi}-\mathrm{Sb}/\mathrm{Co}$ heterostructures},
  author = {Roschewsky, Niklas and Walker, Emily S. and Gowtham, Praveen and Muschinske, Sarah and Hellman, Frances and Bank, Seth R. and Salahuddin, Sayeef},
  journal = {Phys. Rev. B},
  volume = {99},
  issue = {19},
  pages = {195103},
  numpages = {5},
  year = {2019},
  month = {May},
  publisher = {American Physical Society},
  doi = {10.1103/PhysRevB.99.195103},
  url = {https://link.aps.org/doi/10.1103/PhysRevB.99.195103}
}

@article{AHN202312,
title = {Observation of the Nernst effect in a $\mathrm{GeTe}/\mathrm{NiFe}$ structure},
author = {Jeong Ung Ahn and Jeehoon Jeon and Seong Won Cho and OukJae Lee and Suyoun Lee and Hyun Cheol Koo},
journal = {Current Applied Physics},
volume = {49},
pages = {12-17},
year = {2023},
issn = {1567-1739},
doi = {https://doi.org/10.1016/j.cap.2023.02.008},
url = {https://www.sciencedirect.com/science/article/pii/S1567173923000317},
}

@article{PhysRevB87020402,
  title = {Matching domain-wall configuration and spin-orbit torques for efficient domain-wall motion},
  author = {Khvalkovskiy, A. V. and Cros, V. and Apalkov, D. and Nikitin, V. and Krounbi, M. and Zvezdin, K. A. and Anane, A. and Grollier, J. and Fert, A.},
  journal = {Phys. Rev. B},
  volume = {87},
  issue = {2},
  pages = {020402},
  numpages = {5},
  year = {2013},
  month = {Jan},
  publisher = {American Physical Society},
  doi = {10.1103/PhysRevB.87.020402},
  url = {https://link.aps.org/doi/10.1103/PhysRevB.87.020402}
}

@article{PhysRevB92064426,
  title = {Dependence of the efficiency of spin Hall torque on the transparency of $\mathrm{Pt}$/ferromagnetic layer interfaces},
  author = {Pai, Chi-Feng and Ou, Yongxi and Vilela-Le\~ao, Luis Henrique and Ralph, D. C. and Buhrman, R. A.},
  journal = {Phys. Rev. B},
  volume = {92},
  issue = {6},
  pages = {064426},
  numpages = {12},
  year = {2015},
  month = {Aug},
  publisher = {American Physical Society},
  doi = {10.1103/PhysRevB.92.064426},
  url = {https://link.aps.org/doi/10.1103/PhysRevB.92.064426}
}

@article{haney2013current,
  title={Current induced torques and interfacial spin-orbit coupling: Semiclassical modeling},
  author={Haney, Paul M and Lee, Hyun-Woo and Lee, Kyung-Jin and Manchon, Aur{\'e}lien and Stiles, Mark D},
  journal={Physical Review B},
  volume={87},
  number={17},
  pages={174411},
  year={2013},
  publisher={APS}
}

@article{10106315027855,
    author = {Takeuchi, Yutaro and Zhang, Chaoliang and Okada, Atsushi and Sato, Hideo and Fukami, Shunsuke and Ohno, Hideo},
    title = {Spin-orbit torques in high-resistivity-$\mathrm{W}/\mathrm{CoFeB}/\mathrm{MgO}$},
    journal = {Applied Physics Letters},
    volume = {112},
    number = {19},
    pages = {192408},
    year = {2018},
    month = {05},
    issn = {0003-6951},
    doi = {10.1063/1.5027855},
    url = {https://doi.org/10.1063/1.5027855},
}

@article{10106315037391,
    author = {Neumann, Lukas and Meinert, Markus},
    title = {Influence of the Hall-bar geometry on harmonic Hall voltage measurements of spin-orbit torques},
    journal = {AIP Advances},
    volume = {8},
    number = {9},
    pages = {095320},
    year = {2018},
    month = {09},
    issn = {2158-3226},
    doi = {10.1063/1.5037391},
    url = {https://doi.org/10.1063/1.5037391},
}

@article{PhysRevB87144411,
  title = {Theory of spin Hall magnetoresistance},
  author = {Chen, Yan-Ting and Takahashi, Saburo and Nakayama, Hiroyasu and Althammer, Matthias and Goennenwein, Sebastian T. B. and Saitoh, Eiji and Bauer, Gerrit E. W.},
  journal = {Phys. Rev. B},
  volume = {87},
  issue = {14},
  pages = {144411},
  numpages = {9},
  year = {2013},
  month = {Apr},
  publisher = {American Physical Society},
  doi = {10.1103/PhysRevB.87.144411},
  url = {https://link.aps.org/doi/10.1103/PhysRevB.87.144411}
}

@article{cao2019measurement,
  title={Measurement of spin mixing conductance in $Ni_{81}Fe_{19}/\alpha$-W and $Ni_{81}Fe_{19}/\beta$-W heterostructures via ferromagnetic resonance},
  author={Cao, W and Liu, J and Zangiabadi, A and Barmak, K and Bailey, WE},
  journal={Journal of Applied Physics},
  volume={126},
  number={4},
  year={2019},
  publisher={AIP Publishing}
}
\end{document}